\documentclass[12pt]{article}
\usepackage{psfig}

\font\doce=cmr12
\def\argsinh{\protect\hbox{\doce argsinh}}

\begin{document}

\begin{center}
{\Large{\bf PHOTOPRODUCTION OF SCALAR\\
\vspace{0.3cm}

MESONS ON PROTONS AND NUCLEI}}
\end{center}

\vspace{1cm}

\begin{center}
{\large{E. Marco$^{1,2}$, E. Oset$^{1,2}$ and H. Toki$^{1}$}}
\end{center}

\begin{center}
{\small{$^{1}$ \it Research Center for Nuclear Physics,
Osaka University, Ibaraki,\\
 Osaka 567-0047, Japan}}

{\small{$^{2}$ \it Departamento de F\'{\i}sica Te\'orica and IFIC, Centro
Mixto Universidad de\\ Valencia-CSIC,
46100 Burjassot (Valencia), Spain}}
\end{center}

\vspace{3cm}

\begin{abstract}
{\small{We study the photoproduction of scalar mesons close to the threshold
of $f_0(980)$ and $a_0(980)$ using a unitary chiral model.
Peaks for both resonances show up in the invariant mass distributions
of pairs of pseudoscalar mesons. A discussion is made on the photoproduction
of these resonances in nuclei, which can shed light on their nature, a
subject of continuous debate.
}}
\end{abstract}

\vspace{2cm}


\newpage

\section{Introduction}

The understanding of the scalar sector of mesons has been traditionally
very problematic. The low energy scalar states, like the $f_0(980)$
$I^G(J^{PC})=0^+(0^{++})$ and $a_0 (980)$ $I^G(J^{PC})=1^-(0^{++})$,
have been ascribed to conventional $q \bar q$ mesons \cite{Morgan,Tornqvist},
$q^2 {\bar{q}}^2$ states \cite{Jaffe,Achasov}, $K \bar K$ molecules
\cite{Weinstein,Jansen}, glueballs \cite{Jaffe2} and hybrids \cite{Barnes}.
An important step in the understanding of the nature of these states
has been made possible in terms of chiral Lagrangians \cite{GasLeu} by
using a nonperturbative unitary model in coupled channels based
on a $O(p^2)$ expansion of the inverse of the $K$ matrix
\cite{OllOsePel}, similar
to the effective range expansion in Quantum Mechanics. 
Within this method, a good reproduction of all data on meson-meson
interactions up to 1.2 GeV is obtained, including the scalar and vector
resonances, with their position, width and partial decay rates well described.

A further insight into the problem is offered in \cite{OllOse}
where an investigation of the meson-meson data up to
1.4 GeV using arguments based on the large $N_c$ limit of QCD is done.
This allows one to distinguish between meson resonances which survive
in the large $N_c$ limit, which are genuine QCD
meson states (essentially built from $q \bar{q}$), and other states
which appear from multiple scattering of the mesons and which qualify
as quasibound meson-meson states or scattering resonances. The genuine
$q \bar{q}$ states in the scalar sector ($L=0$) would be one octet around
1.4 GeV and a singlet around 1 GeV. The $\sigma(500)$ and $a_0(980)$
appear then as a $\pi \pi$ resonance and a quasibound meson-meson state,
respectively. The $f_0(980)$ becomes a mixture of the genuine 1 GeV
singlet with large components of a meson-meson quasibound state.
This 1 GeV singlet could be associated with the $I=0$
state predicted around this energy in QCD inspired models \cite{Peris,Espriu}
and also has been advocated in phenomenological analyses \cite{Au,MorPen}.
The effects of the 1 GeV singlet are essentially seen in the
$\eta \eta$ decay channels at energies above 1.1 GeV, but it
has no practical effect on other channels. This is indicative of the
large weight of the meson-meson molecular component in the $f_0(980)$
resonance. For this reason it was possible to obtain
a good reproduction of the data of the scalar sector below 1.2 GeV in
terms of multiple scattering of the mesons alone with a ``pseudopotential''
provided by the lowest order chiral Lagrangian \cite{OllOse2}.
In addition, one needs there a cut off
to regularize the loop integrals and account for the effect of
higher order contributions from the second order Lagrangian.
This latter picture is technically very simple and it is the one we shall
employ here.

Consistency of these ideas, and in any case information on
the nature of these states, can be obtained by producing them on proton
targets and also in a
nuclear environment which modifies the properties of the building
blocks, in this case pseudoscalar mesons, and should have repercussions
on the scalar mesons. Hence, we make some suggestions on how
the properties of the $f_0(980)$ and $a_0(980)$
scalar resonances in the nuclear medium could be investigated.

\section{Photoproduction of the scalar resonances}

The reaction proposed is

\begin{equation}
\gamma p \rightarrow p M \, ,
\end{equation}
where $M$ is either of the resonances $f_0$ or $a_0$.
In practice, the meson $M$ will
decay into two mesons, $\pi \pi$ or $K \bar K$ in the case of the $f_0(980)$
or $K\bar K$, $\pi \eta$ in the case of the $a_0(980)$. 

\begin{figure}
\centerline{
\hbox{
\centerline{\protect\hbox{
\psfig{file=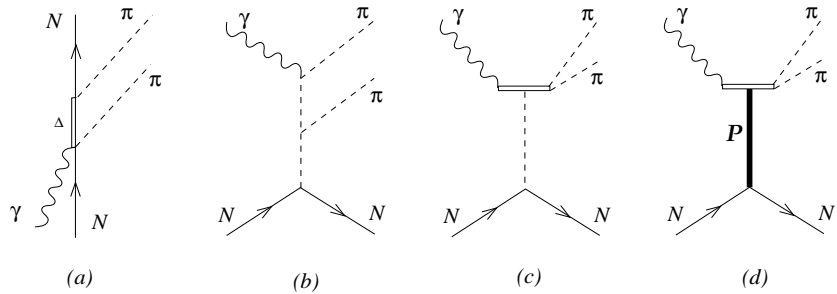,width=0.8\textwidth,silent=,angle=-90}}}
}}
\caption{Mechanisms for two pion production considered in Refs.\ 
\protect\cite{Tejedor, Ochi, Adam,Fries}. 
}
\end{figure}

The $f_0$ and $a_0$ are $L=0$ resonances. Hence we introduce the
basic mechanisms that will lead to the s-wave production of the
pair of mesons.
Photoproduction of pairs of mesons has been the subject of theoretical
studies \cite{Tejedor, Ochi, Adam},
particularly for $\pi \pi$ production. The approaches
of \cite{Tejedor, Ochi} for $\pi \pi$ production
are meant for energies of the photon
below the $K^+ K^-$ production threshold. Without going into detail
in these latter models, we can refer to the dominant term, depicted in Fig.\
1($a$), in order to see that it does not involve the production of the two
pions in s-wave. Indeed, the upper vertex corresponds to
$\Delta \rightarrow \pi N$ decay, where the pion is produced in a p-wave
and the lower vertex is of the type $\vec{S}^+\cdot \vec{\epsilon}$\,
leading to an s-wave pion. Another term of relevance in \cite{Tejedor, Ochi}
is the photoproduction of the $N^\ast$(1520) resonance which later
decays into a $\Delta \pi$ (with s and d-waves) and the $\Delta$ again decays
into a $\pi N$ with the pion in p-wave. Other terms containing explicitly
the production of a $\rho$ meson involve directly the pions in p-wave.

One can think of other resonance excitation like the $N^\ast(1535)$
and its decay into $N\pi\pi$, but this we will show later on that does
not lead to $2\pi$ in s-wave.

A different point of view can be taken by means of which the production
of the s-wave pair of mesons is isolated. This can be accomplished
easily in the context of chiral effective theories which are meant
to work at relatively low energies. This is the approach which we shall
follow and for this reason we shall concentrate at energies of the photon
close to the threshold production of the $f_0(980)$ and $a_0(980)$ 
resonances.

In \cite{Adam} a combined analysis of $\pi\pi$ and $K\bar{K}$
photoproduction in s-wave is conducted. The study is done at higher photon
energies than in the present paper, $E_{\gamma} = 4$ GeV in \cite{Adam}
while here we shall evaluate cross sections for $E_{\gamma} = 1.7$ GeV.
In \cite{Adam} a particular mechanism for pair production is used, which
is depicted in Fig.\ 1($b$). The intermediate meson lines stand for
$\pi$, $\rho$, $\omega$. Alternatively a Regge exchange model is used with
a strength to be adjusted to data.
At the high energies explored and with the model used, there is
a strong $t$ dependence of the cross section and the pair of pions
are produced in many partial waves, out of which the s-wave is projected
out.

In \cite{Fries} which is concerned about $K^+ K^-$ photoproduction,
other mechanisms are suggested. They are depicted in Fig.\ 1($c$),
where an s-wave resonance is produced from the photon and a virtual
meson or a Pomeron (as in diagram 1($d$)) is exchanged from a vector meson 
produced by the photon. In addition, the bremsstrahlung diagrams
depicted in Fig.\ 2$(a)(b)$ are also suggested. It looks clear to us that
at high energies the mechanisms of production can be rather complicated,
as the complexity of the $\phi$ photoproduction model of
\cite{Titov}, followed by $\phi \rightarrow K^+K^-$ decay, shows. Also,
as shown in \cite{Titov}, there are many unknown parameters in the theory.
The same can be said about the diagrams $(c)$ and $(d)$ of Fig.\ 1,
the strength of which is unknown.

Our approach is different to all of these and relies upon the use of effective
chiral Lagrangians. They provide us with Lagrangians for s-wave coupling
of pairs of mesons to the baryons, from where the coupling of the external
photon becomes straightforward. However, we have the limitation of relatively
low energies for the use of these Lagrangians and this is the reason why
we concentrate around threshold of the scalar mesons production.

The lowest order Lagrangians for meson-meson and meson-baryon interactions
are given by \cite{GasLeu}

\begin{equation}
\mathcal{L}_{\mbox{\scriptsize I}} = 
\mathcal{L}_{\mbox{\scriptsize I}}^{\mbox{\scriptsize (M)}} +
\mathcal{L}_{\mbox{\scriptsize I}}^{\mbox{\scriptsize (MBE)}} + 
\mathcal{L}_{\mbox{\scriptsize I}}^{\mbox{\scriptsize (MBO)}}
\end{equation}
with $\mathcal{L}_{\mbox{\scriptsize I}}^{\mbox{\scriptsize (M)}}$
for pure meson-meson interaction,
$\mathcal{L}_{\mbox{\scriptsize I}}^{\mbox{\scriptsize (MBE)}}$ for
meson-baryon vertices containing an even number of mesons and
$\mathcal{L}_{\mbox{\scriptsize I}}^{\mbox{\scriptsize (MBO)}}$ for
meson-baryon vertices containing an odd number of mesons. These
interaction Lagrangians are given by

\begin{equation}                     \label{eq:M}
\mathcal{L}_{\mbox{\scriptsize I}}^{\mbox{\scriptsize (M)}} = 
\frac{1}{12 f^2} \langle (\partial_{\mu} \Phi \Phi -\Phi 
\partial_{\mu} \Phi)^2 + M \Phi^4 \rangle
\end{equation}

\begin{equation}                     \label{eq:MBE}
\mathcal{L}_{\mbox{\scriptsize I}}^{\mbox{\scriptsize (MBE)}} = 
\frac{1}{4 f^2} \langle \bar{B} i \gamma^{\mu} \left[(\Phi \partial_{\mu} \Phi
- \partial_{\mu} \Phi \Phi)B - B(\Phi \partial_{\mu} \Phi
- \partial_{\mu} \Phi \Phi)\right]\rangle
\end{equation}

\begin{equation}                     \label{eq:MBO}
\mathcal{L}_{\mbox{\scriptsize I}}^{\mbox{\scriptsize (MBO)}} = 
\frac{D+F}{2} \langle \bar{B} \gamma^{\mu} \gamma^5 u_{\mu} B \rangle +
\frac{D-F}{2} \langle \bar{B} \gamma^{\mu} \gamma^5 B u_{\mu} \rangle
\end{equation}
with $u_{\mu}$ up to three meson fields given by

\begin{equation}
u_{\mu} = - \frac{\sqrt{2}}{f} \partial_{\mu} \Phi +
\frac{\sqrt{2}}{12 f^3} (\partial_{\mu} \Phi \Phi^2 - 
2 \Phi \partial_{\mu} \Phi \Phi + \Phi^2 \partial_{\mu} \Phi)
\end{equation}
with $f$ the pion decay constant, the symbol $\langle ~\rangle$ standing
for the trace of the $SU(3)$ matrices and $\Phi$, $B$ the meson and baryon
$SU(3)$ matrices given by

\begin{equation}
\Phi = \left( \begin{array}{ccc}
\frac{1}{\sqrt{2}} \pi^0 + \frac{1}{\sqrt{6}} \eta & \pi^+ & K^+\\
\pi^-& -\frac{1}{\sqrt{2}} \pi^0 + \frac{1}{\sqrt{6}} \eta & K^0\\
K^-& \bar{K}^0 &-\frac{2}{\sqrt{6}} \eta
\end{array} \right)
\end{equation}
\begin{equation}
B = \left( \begin{array}{ccc}
\frac{1}{\sqrt{2}} \Sigma^0 + \frac{1}{\sqrt{6}} \Lambda & \Sigma^+ & p\\
\Sigma^-& -\frac{1}{\sqrt{2}} \Sigma^0 + \frac{1}{\sqrt{6}} \Lambda & n\\
\Xi^-& \Xi^0 &-\frac{2}{\sqrt{6}} \Lambda
\end{array} \right)
\end{equation}
Our starting point will be the meson-baryon $\rightarrow$ meson-baryon
vertex originated from the Lagrangian of Eq.\ (\ref{eq:MBE}). We need
actually only the $K^- p \rightarrow K^- p$ and
$\pi^- p \rightarrow \pi^- p$ couplings which are given by

\begin{equation}        \label{eq:vertex}
V_{\pi (K)} = -C_{\pi (K)} \frac{1}{4f^2} \bar u(p') \gamma^\mu
u(p) (k_{\mu} + k'_{\mu}) \, ,
\end{equation}
where $k$, $k'$ are the momenta of the incoming and outgoing
mesons and $C_{\pi}=1$, $C_K =2$. 
As mentioned above, the constant $f$ is the pion decay
constant ($f_{\pi} = 93$ MeV). In the present work the $K \bar K$ states
are the most relevant building blocks of the resonances, and thus the
$K^- p \rightarrow K^- p$ is the relevant ingredient. In
Refs.~\cite{KaiSieWei,OseRam} a study of the latter reaction and coupled
channels was done using a unitary chiral method. We follow here the
approach of \cite{OseRam} where an intermediate
value of $f$ between the one of kaons and pions $f=1.15 f_{\pi}$
was chosen and this will be used here too for this latter amplitude.
The choice of an average value for $f$ in \cite{OseRam},
together with the choice
of a cut off, is an approximate way to incorporate the effects of higher
order Lagrangians, which is possible in the $K^- p$ sector but not in other
meson baryon channels \cite{KaiSieWei, Assum}.

\begin{figure}
\centerline{
\hbox{
\centerline{\protect\hbox{
\psfig{file=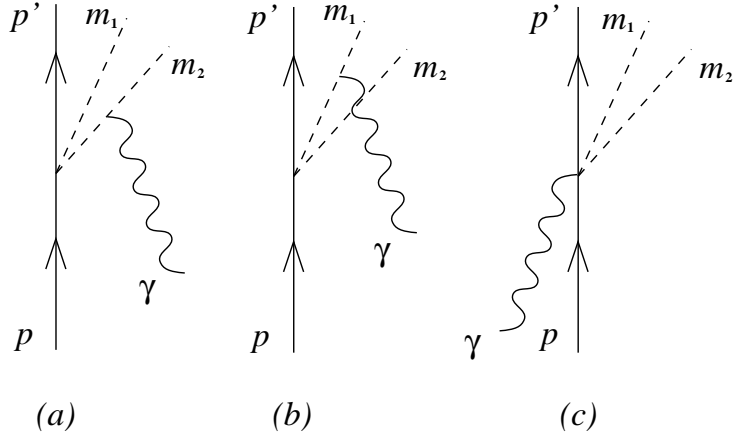,width=0.7\textwidth,silent=,angle=-90}}}
}}
\caption{Feynman diagrams for the process 
$\gamma p \rightarrow p \;m_1 m_2$: $(a)$ and $(b)$, meson
pole processes; $(c)$ contact term required by gauge invariance.}
\end{figure}

For the low energies which we will consider here
and only s-wave of the pair of mesons,
the vertex of Eq.~(\ref{eq:vertex}) simplifies to

\begin{equation}        \label{eq:vertex_simplified}
V_{\pi (K)} = -C_{\pi (K)} \frac{1}{4f^2} (k^0 + k'^0) \, .
\end{equation}
This vertex, together with the standard electromagnetic coupling
of the photon to the mesons, allows one to evaluate diagrams
$(a)$ and $(b)$ of Fig.~2. However, gauge invariance requires
the presence of the contact term of Fig.~2$(c)$, which we also need,
for $\gamma \pi^- p \rightarrow \pi^- p$ and
$\gamma K^- p \rightarrow K^- p$, or analogously
$\gamma p \rightarrow \pi^+ \pi^- p$,
$\gamma p \rightarrow K^+ K^- p$. The vertex is given by

\begin{equation}        \label{eq:elec_vertex}
V^{\gamma}_{\pi (K)} = -C_{\pi (K)} \frac{e}{2f^2} \bar u(p') \gamma^\mu
u(p) \epsilon_{\mu} \, .
\end{equation}
As argued above we choose an energy
of the photon around $E_{\gamma \mbox{\scriptsize lab}} = 1.7$ GeV.
This allows one to produce the scalar resonances close to threshold.
This kinematics allows us to simplify
Eq.~(\ref{eq:elec_vertex}) which becomes now in the CM of the
$\gamma p$ system and using the Coulomb gauge, $\epsilon^0 = 0$,
$\vec{\epsilon} \cdot \vec{k} = 0$,

\begin{equation}        \label{eq:elec_vertex_cm}
V^{\gamma}_{\pi (K)} = C_{\pi (K)} \frac{e}{2f^2}
\frac{i(\vec{\sigma} \times \vec{q}) \vec{\epsilon}}{2M} \, ,
\end{equation}
with $M$ the mass of the proton and $\vec{q}$ the photon momentum.

For the case of $K^+ K^-$ production, with the energy of the photon
chosen, the kaon momenta are very small. In this case, the kaon
Bremsstrahlung diagrams, 2$(a)$, 2$(b)$, give a negligible contribution
(less than 5\%) and we shall neglect them. This is not the case for
the pions, which carry a larger momentum and these
mechanisms become important. On the other hand, there are many
mechanisms for $\pi^+ \pi^-$ production around this region, as can be
seen by the relative complexity of the models used to study the process
$\gamma p \rightarrow \pi^+ \pi^- p$ up to $E_{\gamma} \simeq 1$ GeV in
\cite{Tejedor,Ochi}.

In the case of $\pi \pi$ production we shall evaluate the contribution
from the $f_0$ resonance and we will estimate the background from the
experimental cross section. This should give us an idea of the ratio
of the signal for $f_0$ excitation to the background to be found in
actual experiments. On the other hand, the near threshold cross sections
for $K^+ K^-$ production evaluated here should be rather realistic,
since other terms which can be constructed in analogy to
the model of \cite{Tejedor} would vanish at threshold.

\begin{figure}
\centerline{
\hbox{
\centerline{\protect\hbox{
\psfig{file=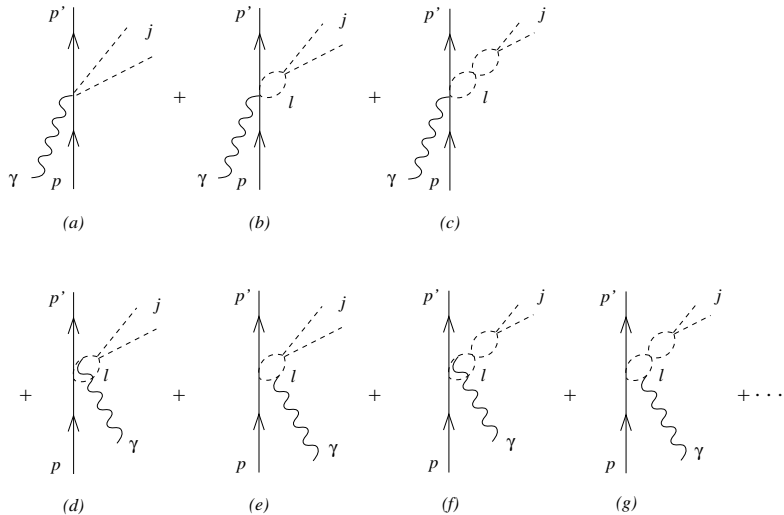,width=0.75\textwidth,silent=,angle=-90}}}
}}
\caption{Iterated terms from the contact term and from mesonic Bremsstrahlung.}
\end{figure}

The next step necessary to build up the scalar meson resonances
is the final state interaction of the mesons. For this purpose we follow
the approach of \cite{OllOse2}, where the resonances are obtained through
an iteration of the lowest order chiral Lagrangian vertex considered
as a potential in the Bethe-Salpeter equation. This is depicted in 
Figs.~3$(a)$, 3$(b)$, 3$(c)$.

However, unlike the tree level Bremsstrahlung diagrams of
Figs.~2$(a)$, 2$(b)$ which are either negligible at threshold of
the meson pair production, or have a strong angular dependence
when the meson momenta are not small, the loops considered in
Figs.~2$(d)$, $(e)$, $(f)$, $(g)$ directly contribute to
s-wave pair production and are also required by gauge invariance.
The set of diagrams in Fig.~3 build up the s-wave resonance
production and are evaluated below.

We have mentioned above how the main terms in $\pi \pi$ production
in \cite{Tejedor, Ochi} do not produce the two pions in s-wave.
One can envisage other mechanisms for the s-wave resonance production
like the one corresponding to diagrams 3 $(a)$, $(b)$, $(c)$, where
the photon couples to the nucleon before of after the $NNMM$ vertex.
In this case the dominant component would vanish at threshold
of resonance production since it involves
the amplitude of Eq.~(\ref{eq:vertex_simplified}) but with
$(k^0 - k'^0)$ rather than $(k^0 + k'^0)$. Smaller components
from $\vec{\gamma} \cdot (\vec{k} - \vec{k}' \;)$ from
Eq.~(\ref{eq:vertex}) would be even further suppressed, since
at threshold of resonance production there is a cancellation
between the diagrams where the photon couples before and after
the $NNMM$ vertex. It is also easy to see that for parity
reasons, terms like those in Figs.~3$(f)$, $(g)$ with the photon
coupling to the second loop do vanish.

Unitarity in coupled channels for the two strongly interacting mesons
is one of the important ingredients here in order to produce the $f_0(980)$
and $a_0(980)$ resonances. One could also think of the coupling of the
final $BMM$ system to intermediate $BM$ states. In this case one should
select the case with $BM$ in s-wave. The coupling of $\gamma N$ to $BM$ in
s-wave has been worked out in \cite{KaiSieWei}
and diagrammatically it is depicted
in Fig.\ 4($a$), incorporating the $BM$ final state interaction.

\begin{figure}
\centerline{
\hbox{
\centerline{\protect\hbox{
\psfig{file=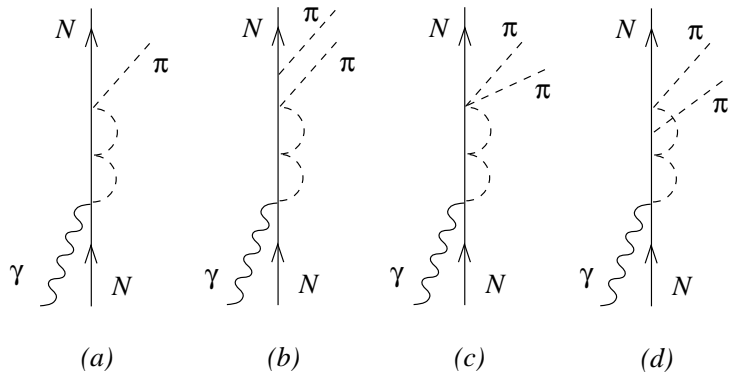,width=0.7\textwidth,silent=,angle=-90}}}
}}
\caption{Diagrams for one and two pion production including coupled channels.
}
\end{figure}

Within the set of chiral Lagrangians written in Eqs.\ 
(\ref{eq:M})--(\ref{eq:MBO}) the way to couple an extra meson on top of the
intermediate $MB$ state is through diagrams like those depicted in Figs.\ 
4$(b)(c)(d)$. All these terms require the use of the 
$\mathcal{L}_{\mbox{\scriptsize I}}^{\mbox{\scriptsize (MBO)}}$ Lagrangian
of Eq.\ (\ref{eq:MBO}), where the $\gamma^{\mu} \gamma_5 \partial_{\mu}
\Phi$ combination leads to a $\vec{\sigma} \vec{p}$
vertex in the nonreativistic reduction, and hence a p-wave meson,
leading to terms which do not contribute to the scalar meson production.

In the particular case of the $N^\ast(1535)$ excitation, the arguments above
can be expressed by stating that the $N^\ast$ with negative parity cannot
decay into a nucleon and two pions in s-wave.

If one goes beyond the chiral approach and considers mechanisms for
two pion production involving the excitation of resonances, we find two
types of diagrams which would provide contribution . One of them is the
$\gamma N \rightarrow N^\ast$ process followed by 
$N^\ast \rightarrow  N \,\,\pi \pi$ ($I=0$, s-wave). The $N^\ast$ should
be a $1/2^+$ state in this case, and restricting ourselves to resonances
below $\sqrt{s} =2000$ MeV, we find the $N^\ast(1440)$ and the $N^\ast(1710)$.
The mechanism mentioned was considered for the case of the $N^\ast(1440)$
in \cite{Tejedor} and found to be relevant only at threshold, but
negligible at higher energies. Here we consider photons at higher
energies than in \cite{Tejedor} and
the $N^\ast(1710)$ has more chances to be relevant. Unfortunately both the
uncertainty in the width , $\Gamma=50$-250 MeV, and the branching ratio
for decay into $N \pi \pi$ ($I=0$,s-wave) , $B=10$-40\%, introduce large
uncertainties in this contribution. On the other hand, the helicity
amplitude for this resonance has also large uncertainties but seems to
be reasonably smaller than in the $N^\ast(1440)$ case \cite{PDG}.

  The second mechanism would correspond to diagrams like in Fig. 1($a$)
with an $N^\ast$ instead of a $\Delta$ in the intermediate baryon. In this case
we should have a $1/2^-$ state to allow for an s-wave pion in the 
$N^\ast \rightarrow N \pi$ decay. Here we would have the $N^\ast(1535)$
and $N^\ast (1650)$ resonances. The uncertainties here stem from the contact
vertex gamma $N N^\ast \pi$, which from
minimal coupling from the leading constant $N N^\ast \pi$ vertex would be zero,
and in practice should be relatively small.

  The presence of such mechanisms would introduce some elements of
uncertainty in the cross sections evaluated here, although from the
arguments used above these terms do not seem to be large. There is still
another argument that would favour the mechanism chosen here.  Indeed,
all the resonant terms discussed above would provide the maximum
contribution when their propagators are placed on shell in the diagrams,
providing an imaginary part of the amplitude. As we shall see later on,
the mechanisms considered here lead to a peak in the real part of the
amplitude when the $f_0(980)$ resonance is excited which can interfere with
the largely real amplitude of the whole 
$\gamma N \rightarrow \pi \pi N$ process, hence
magnifying the effect of the resonance. This would not be the case of
the $N^\ast$ resonance excitation mechanisms
which would contribute mostly an imaginary part to the amplitude in
the case when the resonance is placed on shell and the mechanism is most
relevant.

  Accepting some uncertainties, the arguments given above indicate that
the mechanism considered here can provide a fair estimate of the
strength of the scalar resonance excitation, which is sufficient for the
purpose of the present paper, where an exploration of the possibilities
of observation of these resonances in gamma induced reactions is made.

Coming back to our model, the final states 
with a pair of mesons which can be produced in
the reactions are $\pi^+ \pi^-$, $\pi^0 \pi^0$, $K^+ K^-$,
$K^0 \bar{K}^0$, $\pi^0 \eta$. Note that even if the $\pi^0 \pi^0$,
$K^0 \bar{K}^0$ and $\pi^0 \eta$ do not couple to the photon vertex
in diagram $(c)$ of Fig.~3, they can appear in the final states through
the iterated terms of diagrams $(b)$, $(c) \ldots (g)$ when we sum over
the intermediate state, $l$, which can be either $\pi^+ \pi^-$
or $K^+ K^-$.

The sum of diagrams in Fig.~3 bears a close resemblance to the $\phi$ decay
into $K \bar{K} \gamma $  which has been studied in
\cite{Lucio,CloIsgKum,Ollerplb}. Indeed, the vertex of
Eq.~(\ref{eq:vertex})  changes
$k_{\mu} + k'_{\mu}$ by $k_{\mu} - k'_{\mu}$
when the two mesons are outgoing and has the same structure
as the $\phi \rightarrow K \bar{K}$ vertex which goes as
$\epsilon^{\mu}(\phi) (k_{\mu} - k'_{\mu})$. The results of
\cite{Lucio,CloIsgKum,Ollerplb} are very useful. Using arguments of
gauge invariance it is found there that the sum of the loops in
Figs.~3$(b)$, 3$(d)$, 3$(e)$ is convergent and replaces the two meson
loop of Fig.~3$(b)$

\begin{equation}
G(P) = i \int \frac{d^4k}{(2\pi)^4}
\frac{1}{k^2 - m^2 +i\epsilon} \frac{1}{(P-k)^2 - m^2 +i\epsilon}
\end{equation}
by
\begin{equation}        \label{eq:Gtilde}
\tilde{G}(Q,P) = - \frac{Q \cdot q}{(2 \pi)^2} 
\int_0^1 dx \; \int_0^x dy \;
\frac{(1-x)y}{Q^2x(1-x) - 2 Q\cdot q (1-x)y - m^2 +i\epsilon}
\end{equation}
where $P$ is the fourmomentum of the two mesons, $m$ the mass
of the meson in the loop ($\pi^+$ or $K^+$ in our case)
and $Q = p - p'$. In addition, $M_I$ is the invariant mass of the pair
of mesons and the invariant product $Q \cdot q$ is given here by 

$$
Q \cdot q = \frac{1}{2} (Q^2 - M_I^2)
$$

\begin{equation}
Q^2 = 2 M^2 - 2 E(\vec{p})E(\vec{p}\; ') + 2|\vec{p}||\vec{p}\; '| \cos \theta
\end{equation}
This introduces a dependence of the $t$ matrix for the process in
the angle of $\vec{p}$, $\vec{p} \; '$, but not the angle of the mesons
with the photon.

\begin{figure}
\centerline{
\hbox{
\centerline{\protect\hbox{
\psfig{file=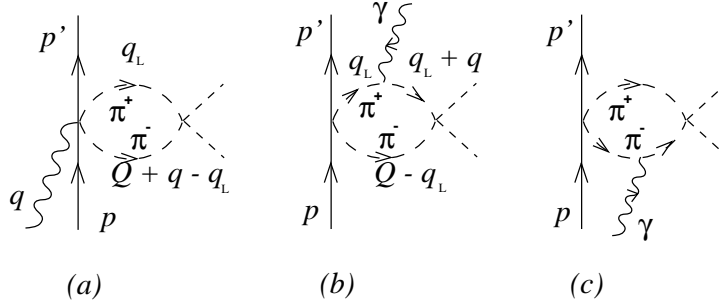,width=0.7\textwidth,silent=,angle=-90}}}
}}
\caption{One loop diagrams for the process $\gamma p \rightarrow p M M$.
}
\end{figure}

The evaluation proceeds as follows. We must evaluate the contribution of the
three loops which we show now in Fig.\ 5 with the appropriate
labels for the momenta of the particles.

Diagrams $(b)$ and $(c)$ contribute on equal amount. We shall call contact
term (C) the one coming from diagram $(a)$ and Bremsstrahlung (B) the one
coming from diagrams $(b)$, $(c)$. The contribution goes as (for
the case of intermediate $\pi^+\pi^-$)

\begin{eqnarray}             \label{eq:tgamC}
-it^{(\gamma \mbox{\scriptsize C})} &=& -C_{\pi} \frac{e}{2f^2}
\bar{u}(\vec{p}\,') \gamma^{\mu} u(\vec{p}) \epsilon_{\mu}
\int \frac{d^4 q_L}{(2\pi)^4}\nonumber \\
&&\times \frac{1}{q_L^2 - \mu^2 +i\epsilon}
\frac{1}{(Q+q-q_L)^2 - \mu^2 +i\epsilon}
t^{(S)}
\end{eqnarray}

\begin{eqnarray}             \label{eq:tgamB}
-it^{(\gamma \mbox{\scriptsize B})} &=& C_{\pi} \frac{e}{2f^2}
\bar{u}(\vec{p}\,') \gamma^{\mu} u(\vec{p})
\int \frac{d^4 q_L}{(2\pi)^4} 2eq_L^{\nu} \epsilon_{\nu}
(2q_L -Q)_{\mu} \nonumber\\
&&\times \frac{1}{q_L^2 - \mu^2 +i\epsilon}
\frac{1}{(Q-q_L)^2 - \mu^2 +i\epsilon}
\frac{1}{(q_L+q)^2 - \mu^2 +i\epsilon}
t^{(S)}
\end{eqnarray}
where $t^{(S)}$ is the strong meson-meson amplitude. Since $Q$, $q$
are the only vectors not integrated in Eqs.\ (\ref{eq:tgamC}) ,
(\ref{eq:tgamB}), the sum of the two terms has a structure of the type

\begin{equation}            \label{eq:abcde}
\gamma^{\mu} \epsilon^{\nu} \left\{
a\, g_{\mu \nu} + b\, Q_{\mu} Q_{\nu} + c\, Q_{\mu} q_{\nu} + 
d\, Q_{\nu} q_{\mu} + e\, q_{\mu} q_{\nu} \right\}\, ,
\end{equation}
where the contact term only contributes to $a g_{\mu \nu}$ while the
B term contributes to all.

Gauge invariance of the sum of all terms requires that the expression
of Eq.\ (\ref{eq:abcde}) vanishes with the substitution
$\epsilon^{\nu} \rightarrow q^{\nu}$. This implies

\begin{equation}
a\, q^{\mu} + b\, Q^{\mu} (Q\cdot q) + d\, q^{\mu} (Q\cdot q) = 0
\end{equation}
or equivalently

\begin{equation}            \label{eq:ba}
b = 0; \quad a = -d (Q\cdot q)\, .
\end{equation}
On the other hand, in the Coulomb gauge chosen where $\epsilon^0 = 0$,
$\vec{\epsilon} \, \vec{q} = 0$ in the $\gamma p$ CM frame, the expression
of Eq.\ (\ref{eq:abcde}) is greatly simplified since we have that

\begin{equation}
\epsilon^{\mu} Q_{\mu} = - \vec{\epsilon} \,\vec{Q} =
- \vec{\epsilon} \, (\vec{p} - \vec{p}\,') \simeq - \vec{\epsilon}\, \vec{p} =
\vec{\epsilon}\, \vec{q} = 0\,,
\end{equation}
\begin{equation}
\epsilon^{\mu} q_{\mu} = - \vec{\epsilon} \,\vec{q} = 0\,,
\end{equation}
where we have assumed $\vec{p}\,' \simeq0$ because we work close to
the scalar meson photoproduction threshold. Hence only the $g_{\mu \nu}$
term of Eq.\ (\ref{eq:abcde}) contributes to the amplitude. The trick
then is to evaluate the term which goes like $d Q_{\nu} q_{\mu}$ to
which only the B diagrams contribute and
from it via Eq.\ (\ref{eq:ba}) obtain the coefficient $a$ which is the
only one needed to evaluate the amplitudes. Eq.\ (\ref{eq:tgamB}) is
then evaluated using the Feynman technique and the terms proportional
to $Q_{\nu} q_{\mu}$ are kept. For dimensional reasons the rest
of the integral has two powers less in the loop variable $q_L$,
and hence the contribution to this term is convergent and via Eq.\
(\ref{eq:ba}) leads to the result of Eq.\ (\ref{eq:Gtilde}).
Further details on the integration technique can be seen
in \cite{Lucio,CloIsgKum,Ollerplb}.

The work of \cite{Ollerplb} adds some new relevant ingredients to the
work of \cite{Lucio,CloIsgKum}, since it proves that by using chiral
Lagrangians to deal with the final state interaction of the mesons
after the first loop which involves the photon, the strong $t$ matrix
for meson-meson interaction factorizes on shell.
This occurs because the $MM\rightarrow MM$ amplitudes in lowest order
have the structure $\alpha s + \beta \sum_i p_i^2$, which can be
recast into $\alpha s + \beta \sum_i m_i^2 + \beta \sum_i (p_i^2-m_i^2)$,
where the first two terms account for the on-shell part. The last term in this
former expression kills one of the meson propagators in Eqs.\ 
(\ref{eq:tgamC}), (\ref{eq:tgamB}) and does not provide contribution
to the $Q_{\nu} q_{\mu}$ term.

With all these ingredients we can write the sum of the diagrams in
Fig.~3 which leads to the amplitude

\begin{equation}        \label{eq:tgamma}
t^{\gamma}_j = \frac{e}{4 f^2} \frac{i (\vec{\sigma} \times \vec{q})
\vec{\epsilon}}{2M} (D_j + \sum_l D_l \tilde{G_l} T_{lj}) \, ,
\end{equation}
where $D_j$ is the vector $(4,0,2,0,0)$ counting the channels in the
following order, $K^+ K^-$, $K^0 \bar{K}^0$, $\pi^+ \pi^-$, $\pi^0
\pi^0$, $\pi^0 \eta$.
The matrix $T_{lj}$ in Eq.~(\ref{eq:tgamma}) is the transition $t$
matrix from the meson state $l$ to $j$. These matrix elements are
easily obtained from \cite{OllOse2} using the isospin decomposition of
the states and we find the matrix of Table~1, in terms of the isospin
$I=0$, $I=1$ matrix elements derived in \cite{OllOse2}.

\begin{table}
\begin{center}
\begin{tiny}
\begin{tabular}{|c|ccccc|}
\hline
&&&&&\\ & $K^+ K^-$ & $K^0 \bar{K}^0$ & $\pi^+ \pi^-$ & $\pi^0
\pi^0$&$\pi^0 \eta$\\ &&&&&\\
\hline
&&&&&\\ $K^+ K^-$ & $\frac{1}{2} \left\{ t_{K \bar{K}, K
\bar{K}}^{I=0} + t_{K \bar{K}, K \bar{K}}^{I=1} \right\}$ &
$\frac{1}{2} \left\{ t_{K \bar{K}, K \bar{K}}^{I=0} - t_{K \bar{K}, K
\bar{K}}^{I=1} \right\}$ & $\sqrt{\frac{1}{3}} t_{K \bar{K}, \pi
\pi}^{I=0} $ & $\sqrt{\frac{1}{3}} t_{K \bar{K}, \pi \pi}^{I=0}$ &
$-\sqrt{\frac{1}{2}} t_{K \bar{K}, \pi^0 \eta}^{I=1}$
\\
&&&& &
\\
$K^0 \bar{K}^0$ & & $\frac{1}{2} \left\{ t_{K \bar{K}, K
\bar{K}}^{I=0} + t_{K \bar{K}, K \bar{K}}^{I=1} \right\} $ &
$\sqrt{\frac{1}{3}} t_{K \bar{K}, \pi \pi}^{I=0} $ &
$\sqrt{\frac{1}{3}} t_{K \bar{K}, \pi \pi}^{I=0} $ &
$\sqrt{\frac{1}{2}} t_{K \bar{K}, \pi^0 \eta}^{I=1}$
\\&&
& &&\\ $\pi^+ \pi^-$ & & & $\frac{2}{3} t_{\pi \pi, \pi \pi}^{I=0} $ &
$\frac{2}{3} t_{\pi \pi, \pi \pi}^{I=0} $ & 0
\\
&& & &&\\ $\pi^0 \pi^0$ & & & & $\frac{2}{3} t_{\pi \pi, \pi
\pi}^{I=0} $ & 0
\\
&& & &&\\ $\pi^0 \eta$ & & & & & $t_{\pi^0 \eta, \pi^0 \eta}^{I=1}$
\\
&& & &&\\
\hline
\end{tabular}
\end{tiny}
\end{center}
\caption{Elements of the transition $t$ matrix from the state $l$ to $j$
$(T_{lj} = T_{jl})$.  The isospin $I=0$, $I=1$ matrix elements can be
found in
\protect\cite{OllOse2}.}
\end{table}

One should note that the matrix elements involving pions use a unitary
normalization in \cite{OllOse2} including an extra factor $1/\sqrt{2}$
per each pair of pion states. This normalization is convenient to
account for factors due to the identity of the particles when summing
over intermediate states. The amplitudes of Table~1 are the physical
ones, where the proper normalization of the states is used.

The resonance structure of the pair of mesons comes from the term $\sum_l
D_l \tilde{G_l} T_{lj}$ in Eq.~(\ref{eq:tgamma}). Hence, for the case of
pion pair production, we remove the isolated term $D_j$ in
Eq.~(\ref{eq:tgamma}) which, together with other terms will build up
the background for this process. In the case of $K \bar{K}$ production
the threshold is above the $f_0$ and $a_0$ mass and the cross section
does not exhibit the resonance structure, although the amplitudes are
affected by it. In this case we keep all terms since with the
amplitude of Eq.~(\ref{eq:tgamma}) we are producing absolute cross
sections.

The function $\tilde{G}$ of Eq.~(\ref{eq:Gtilde}) can be written in an
analytical form following \cite{Lucio,CloIsgKum}. However, there are
novel ingredients here since $Q^2$ can be negative, unlike the case of
the $\phi$ decay, where $m_{\phi}^2$ is positive. For this reason we
give below the analytical expressions valid in all the range of values
of $Q^2$, $M_I^2$.

\begin{eqnarray}
\tilde{G}(Q^2,M_I^2)& =& \frac{1}{8\pi^2} \left\{
\frac{1}{2} - \frac{2}{a-b}\left[f\left(\frac{1}{b}\right) -
f\left(\frac{1}{a}\right)\right]\right. \nonumber \\
&&+
\left.\frac{a}{a-b} \left[g\left(\frac{1}{b}\right) -
g\left(\frac{1}{a}\right)\right]\right\}
\end{eqnarray}
where $a = Q^2/m^2$, $b = M_I^2/m^2$, $m$
is the mass of the meson in the loop and $f(x)$ and $g(x)$ are given by

$$
f(x) = \left\{
\begin{array}{ll}
 -\left[\arcsin\left(\frac{1}{2\sqrt{x}}\right)\right]^2 
&\mbox{for} \,\, x>\frac{1}{4}\\
\\
\frac{1}{4} \left[\ln\left(\frac{\eta_+}{\eta_-}\right) - i \pi\right]^2
&\mbox{for} \,\, 0<x<\frac{1}{4}\\
\\
\left[\argsinh\left(\frac{1}{2\sqrt{-x}}\right)\right]^2
&\mbox{for} \,\, x<0
\end{array}
\right.
$$
\vspace{.5cm}
$$
g(x) = \left\{
\begin{array}{ll}
(4x-1)^{\frac{1}{2}} \arcsin\left(\frac{1}{2\sqrt{x}}\right)
&\mbox{for} \,\, x>\frac{1}{4}\\
\\
\frac{1}{2} (1-4x)^{\frac{1}{2}} \left[\ln\left(\frac{\eta_+}{\eta_-}\right)
- i \pi\right] 
& \mbox{for} \,\, 0<x<\frac{1}{4}\\
\\
(1-4x)^{\frac{1}{2}} \argsinh\left(\frac{1}{2\sqrt{-x}}\right)
& \mbox{for} \,\, x<0
\end{array}
\right.
$$
\vspace{.5cm}
\begin{equation}
\eta_{\pm} = \frac{1}{2x} \left[1 \pm (1-4x)^{\frac{1}{2}}\right]
\end{equation}

The particular structure of Eq.~(\ref{eq:tgamma}) allows one
to obtain an easy formula for the invariant mass distribution of
the two mesons

\begin{eqnarray}        \label{eq:dsigma_final}
\left. \frac{d \sigma}{d M_I} \right|_j &=& \frac{1}{(2\pi)^3}
\frac{1}{4s} \frac{M^2}{s-M^2} \frac{1}{M_I} S
\lambda^{1/2} (s, M^2_I, M^2) \lambda^{1/2} (M_I^2, m_1^2, m_2^2) \nonumber \\
&&\times \frac{1}{2} \int_{-1}^{1} d \cos \theta 
\bar{\sum} \sum \left| t^{\gamma}_j \right|^2\, ,
\end{eqnarray}
where $m_1$, $m_2$ are the masses of the two mesons in the final
meson-meson state, $\lambda$ is the ordinary K\"allen function
and $S$ is a symmetry factor, $1/2$ for $\pi^0 \pi^0$ in the final
state and 1 for the other channels.

\begin{figure}
\centerline{
\hbox{
\centerline{\protect\hbox{
\psfig{file=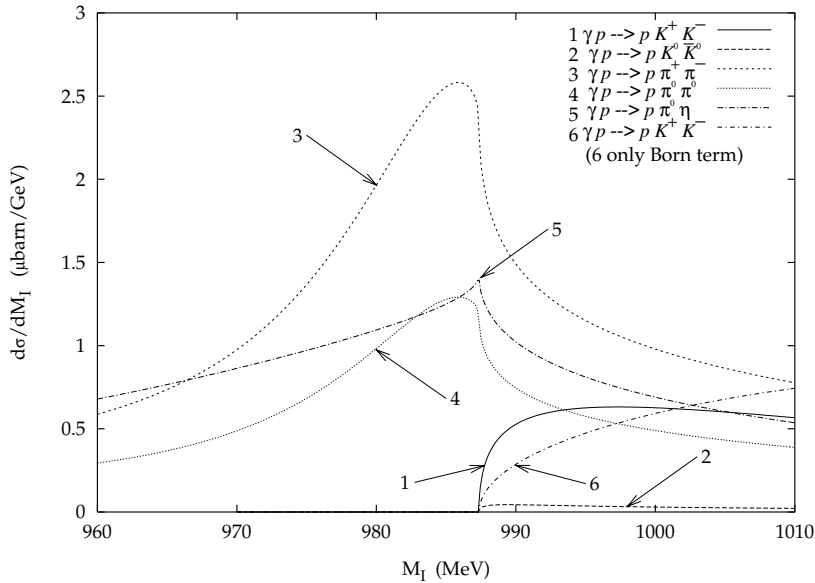,width=0.8\textwidth,silent=,angle=-90}}}
}}
\caption{Results for the cross section on protons
as a function of the invariant
mass of the meson-meson system.}
\end{figure}

The technique used here has been recently applied to the study of the
radiative $\phi$ decay, $\phi \rightarrow \pi^+ \pi^- \gamma$,
which proceeds via $K^+ K^-$ loops \cite{rhophi}. An invariant mass
distribution is predicted in \cite{rhophi} with a clear peak for the
$f_0(980)$ excitation. Recently the measurements have been concluded
at Novosibirsk \cite{Novo} and the experimental distribution is in perfect
agreement with the predictions of \cite{rhophi}. This finding gives us extra
confidence in the techniques used here for the photoproduction processes.

\section{Results}

In Fig.~6 we show the results for the 5 channels considered. We observe
clear peaks for $\pi^+ \pi^-$, $\pi^0 \pi^0$ and $\pi^0 \eta$
production around 980 MeV. The
peaks in $\pi^+ \pi^-$ and $\pi^0 \pi^0$ clearly correspond to the
formation of the $f_0(980)$ resonance, while the one in
$\pi^0 \eta$ corresponds to the formation of the
$a_0(980)$. The $\pi^0 \pi^0$ cross section is $\frac{1}{2}$
of the $\pi^+ \pi^-$ one due to the symmetry factor $S$ in
Eq.~(\ref{eq:dsigma_final}). As commented above,
the $K^+ K^-$ and $K^0 \bar{K}^0$ production cross section appears
at energies higher than that of the resonances and hence do not show
the resonance structure. Yet, final state interaction is very important
and increases appreciably the $K^+ K^-$ production cross section
for values close to threshold with respect to the Born approximation
(only the $D_j$ term in Eq.~(\ref{eq:tgamma}), or diagram $(a)$
of Fig.~3).

It is also interesting to see the shapes
of the resonances which differ appreciably from a Breit-Wigner, due
to the opening of the $K \bar{K}$ channel just above the
resonance \cite{Flatte}.

We would like to stress here that the invariant
mass distributions for resonance excitation into the various
pseudoscalar channels depicted in Fig.~6 are theoretical predictions
of a chiral unitary model, in this case the one of \cite{OllOse2},
where only one parameter was fitted to reproduce all the data of
the meson-meson interaction in the scalar sector.

A small variant of this reaction would be the 
$\gamma p \rightarrow n M \bar{M}$. In this case the $M \bar{M}$
system has charge $+1$ and hence $I=0$ is excluded, hence,
one isolates the $a_0$ production.

\begin{figure}
\centerline{
\hbox{
\centerline{\protect\hbox{
\psfig{file=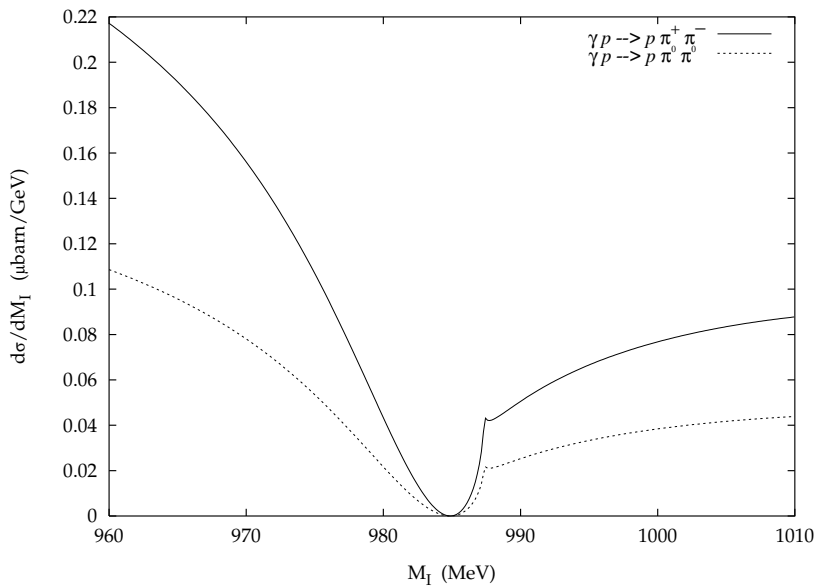,width=0.8\textwidth,silent=,angle=-90}}}
}}
\caption{Results for the cross section on protons
as a function of the invariant
mass of the meson-meson system including only the $\pi^+ \pi^-$ contribution
on the first loop.}
\end{figure}

It is interesting to notice the origin of the peak structure for
$\pi \pi$ production. Indeed, the cross section for
$\pi \pi \rightarrow \pi \pi$ in $I=0$ exhibits a minimum at the
$f_0$ energy because of the interference between the $f_0$
contribution and the $\sigma(500)$ broad resonance. We can see this
here also by killing the $K^+ K^-$ first loop in the diagrams of
Fig.~3. The results are shown
in Fig.~7 for $\pi^+ \pi^-$ and $\pi^0 \pi^0$ production. Very small
cross sections and a clear minimum around the $f_0$ position
can be seen in the figure. This means that the resonant structure for 
$\pi\pi$ production of Fig.~6 is due to the $K^+ K^-$ first
loop which factorizes the $K^+ K^- \rightarrow \pi\pi$ amplitude in the
final state interaction. This amplitude has a peak at the $f_0$ position but
cannot be seen in the $K^+ K^- \rightarrow \pi\pi$ cross section
because the resonance is below threshold. A reaction like the
present one which factorizes this amplitude at energies below the physical
threshold can then evidence the peak, as is indeed visible in the figure.

However, we should bear in mind that we
have plotted there the contribution of the $f_0$ resonance alone. The tree
level contact term and Bremsstrahlung diagrams, plus other contributions
which would produce a background, are not considered there. We estimate
the background from the experimental cross section for
$\gamma p \rightarrow p \pi^+ \pi^-$ of \cite{ABBHHM}, which is
around 45 $\mu b$ at $E_{\gamma} = 1.7$ GeV. Using
Eq.~(\ref{eq:dsigma_final}), assuming $\bar{\sum} \sum |
t_j^{\gamma} | ^2$ constant and integrating over the range
of $M_I$ allowed, we determine that constant from the experimental cross
section and then the same 
Eq.~(\ref{eq:dsigma_final}) gives us the background for $d\sigma
/dM_I$. This provides a background of around 55 $\mu b /$GeV while
the resonant peak has about 2.5 $\mu b/$GeV strength. This gives a ratio
of 5\% signal to background assuming that the
background is mostly real versus an imaginary contribution
from the resonance and hence there would be no interference.
The situation with the $\pi^0 \pi^0$ channel should be better
because the $\gamma p \rightarrow \pi^0 \pi^0 p$ cross section
is about eight times smaller than the one for
$\gamma p \rightarrow \pi^+ \pi^- p$ \cite{Zabrodin,Harter}.
Considering that the resonant signal now is a factor two smaller than the
$\gamma p \rightarrow \pi^+ \pi^- p$ cross section, this would give
a ratio of signal to background of 20\%, which should be more clearly
visible in the experiment. The same or even better ratios than in
the $\pi^0 \pi^0$ case are expected
for $\pi^0 \eta$ production in the $a_0$ channel, since estimates
of the background along the lines of present models for
$\pi^0 \pi^0$ production \cite{Tejedor,Ochi} would provide a
cross section smaller than for $\pi^0 \pi^0$ production. 

\begin{figure}
\centerline{
\hbox{
\centerline{\protect\hbox{
\psfig{file=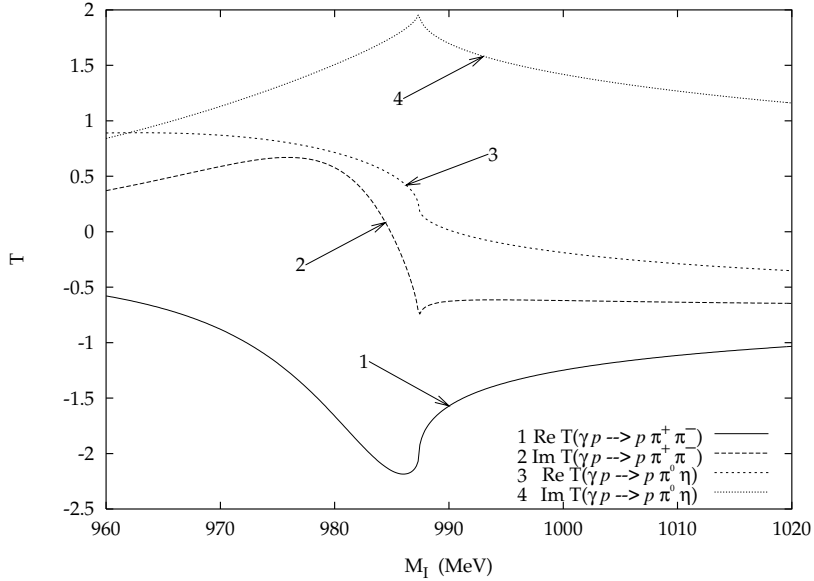,width=0.8\textwidth,silent=,angle=-90}}}
}}
\caption{Real and imaginary parts of the resonant piece of the
amplitude $t^{\gamma}_j$ ($\sum_l D_l \tilde{G_l} T_{lj}$ 
of Eq.~(\protect{\ref{eq:tgamma}})),
for the channels $\pi^+ \pi^-$ and $\pi^0 \eta$.
}
\end{figure}

However, there is a distinct feature about the $f_0$ resonance which
makes its contribution, in principle, bigger than the estimates
given above. Indeed, the $f_0$ is approximately a Breit-Wigner resonance
with an extra phase of $e^{i \pi/2}$. This means that the real part
has a peak while the imaginary part changes sign around the
resonance energy. This was the case in the $K \bar{K} \rightarrow \pi \pi$
amplitude \cite{OllOse3} and we have the same situation here as can be
seen in Fig.~8. This means that assuming the background basically real,
there would be an interference with the $f_0$ resonance which would
lead to an increase of about 50\% over the background, or a decrease
by about 40\% (depending on the relative sign) for the $\pi^+ \pi^-$
case and larger effects for the $\pi^0 \pi^0$ case. This is of course
assuming weak dependence on momenta and
spin of the background amplitudes. In any case, due to the particular
feature of the $f_0$ resonance discussed above, it is quite
reasonable to expect bigger signals than the estimates based on a
pure incoherent sum of cross sections.

Certainly it is possible to obtain better ratios if one looks at angular
correlations. If one looks in a frame where the two mesons are
in their CM, the Bremsstrahlung pieces (both from the squared of
the Bremsstrahlung term as well as from interference with
s-wave terms) have a $\sin^2 \theta$ dependence,
with $\theta$ the angle between the meson and the photon.
Other terms from \cite{Tejedor,Ochi} exhibit equally strong
angular dependence, for what extraction of the angle independent part
of the cross section would 
be an interesting exercise which would select the part
of the cross section to which the resonant contribution
obtained here belongs to.

\section{Meson resonance production in nuclei}

Now we turn our attention to nuclei. 
As mentioned in the introduction there is much debate about the nature of the
scalar meson resonances. The modification of the properties of these
resonances in nuclei should depend on their nature. For instance,
it would not be modified in the same way 
if it is a $q \bar{q}$ state than if it is a 
$K \bar{K}$ molecule. Also, our scheme does not rely upon any of these
pictures, although it gives some support to the quasimolecular nature
of the states. In any case, we saw that the loop structure, with mostly
$K \bar{K}$ in the loop, is what leads to the $f_0$ production. Thus, the
production in nuclei would be modified due to the $K$, $\bar{K}$
modification in a nuclear medium, but in a particular way, due to the
modification of the $\tilde G$ function in a medium when the
$K$, $\bar{K}$ propagators are substituted by their renormalized ones in the
medium. The changes expected would certainly differ form those expected
on the base of the assumption of a $K \bar{K}$ molecule and particularly
a $q \bar{q}$ state for this resonance. In this sense modifications of the
resonance properties in nuclei are bound to offer us some information
on the nature of the states, eventually reinforcing the chiral unitary
approach interpretation of those states. The evaluation of the nuclear
modifications would require the use $K$, $\bar{K}$ selfenergies in
the nuclear medium, or equivalently their optical potentials. 

The interaction of $K$, $\bar{K}$ with nuclei is a subject that has attracted
much attention \cite{todos}. Interesting developments
have been done recently looking at
$K^- N$ scattering from a chiral perspective, which have
allowed to tackle the problem of the $K$, $\bar{K}$ nucleus interaction
with some novel results \cite{Waas, Lutz, RamOsePre}.
The issue is not yet settled since there
are still important discrepancies between the different results. It
looks wise to allow some time for the issue to get clarified before one
tackles the problem suggested here, which looks certainly quite interesting.
Meanwhile we can make some exploration following the lines of the former
sections.
The first thing which we
observe is that if one looks for a proton in the final state,
one can have the
$\gamma n \rightarrow p \pi^- \eta (K^- K^0)$ and approximately
one would expect a cross section

\begin{equation}
\left.\frac{d\sigma}{dM_I}\right|_A \simeq Z\frac{d\sigma}{dM_I} (p)
+ N\frac{d\sigma}{dM_I} (n) \, .
\end{equation}
The latter cross section can proceed through the
meson channels $K^- K^0$ and $\pi^- \eta$, both in $I=1$. The
cross sections in this case, for these two channels and in this
order, are given again by means of Eqs.~(\ref{eq:tgamma})
and (\ref{eq:dsigma_final}), by taking in Eq.~(\ref{eq:tgamma}) the
vector $D=(1,0)$ and the matrix $T_{lj}$ as

\begin{equation}
T_{lj} = \left(
\begin{array}{c c}
t_{K \bar{K}, K \bar{K}}^{I=1} & t_{K \bar{K}, \pi \eta}^{I=1} \\
t_{K \bar{K}, \pi \eta}^{I=1}& t_{\pi \eta, \pi \eta}^{I=1}
\end{array}
\right)\, .
\end{equation}
These cross sections are about one order of magnitude smaller
than those on the proton target, as seen in Fig.~9.

\begin{figure}
\centerline{
\hbox{
\centerline{\protect\hbox{
\psfig{file=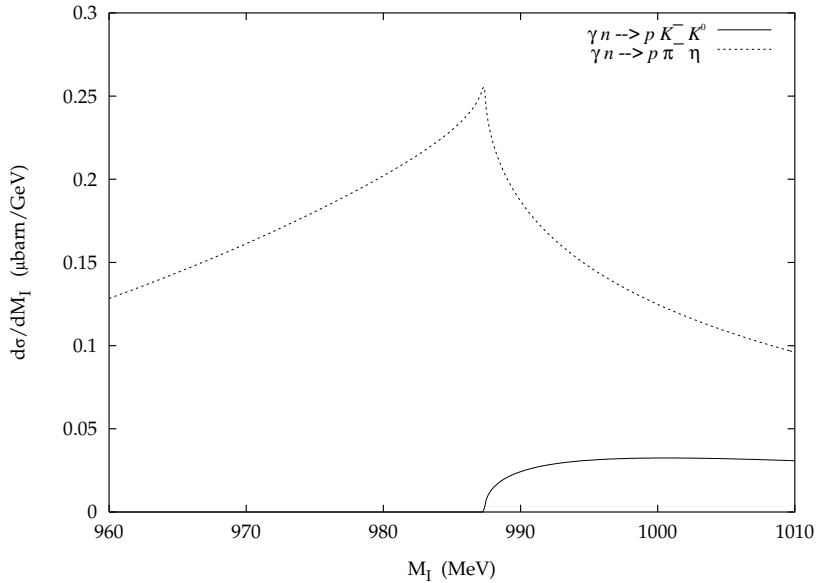,width=0.8\textwidth,silent=,angle=-90}}}
}}
\caption{Results for the cross section on neutrons
as a function of the invariant
mass of the meson-meson system.}
\end{figure}

Hence, in nuclei we should expect a cross section roughly
$Z$ times the one of the proton, unless the properties of
the resonances $a_0$ and $f_0$ are drastically modified
in the medium, which is, however, what one expects. One
should recall that the relatively small widths of the $f_0$ and
$a_0$ resonances are due to the small coupling to the
$\pi \pi$ and $\pi \eta$ channels respectively. The resonances,
however, couple very strongly to the $K \bar{K}$ system but the decay
is largely inhibited because the $K \bar{K}$ threshold
is above the resonance mass. Only the fact that the resonances have
already a width for $\pi \pi$ and $\pi \eta$ decay, respectively,
allows the $K \bar{K}$ decay through the tail of the
resonance distribution. If the $K^-$ develops a large width on
its own this enlarges considerably the phase space for $K \bar{K}$
decay and the $a_0$, $f_0$ width should become considerably larger.

Given the interest that the
modifications of meson resonances in nuclei, like the $\sigma$
\cite{Schuck,Aouissat}, $\rho$ \cite{Soyeur,Hermann}, etc., is
raising, the study of the
modifications of the $f_0$ and $a_0$ is bound to offer us
some insight into the nature of these resonances, that has
been so much debated, and eventually into the chiral approach to these
resonances which we have discussed in this paper.

\section{Conclusions}
In summary, we have studied the photoproduction of the $f_0(980)$ and
$a_0(980)$ resonances for photon energies close to the $K \bar{K}$
production threshold using tools of chiral unitary theory. The $K \bar{K}$
production cross sections were evaluated and the effect of the resonances
was shown to modify drastically the cross sections with respect to the Born
approximation. The $f_0$ and $a_0$ resonances led to peaks in the invariant
mass distributions of $\pi \pi$ and $\pi \eta$ production. Although
large backgrounds are expected, the signals could be visible
particularly if angular correlations are also studied. The
$(\gamma, p)$ experiment in nuclei would also lead to the $f_0$
and $a_0$ excitation mostly from the collision of photons with
protons, since the neutrons provided only a contribution of about
an order of magnitude smaller than protons for $a_0$ production
and do not contribute to the $f_0$ production. The studies in nuclei
would provide information on the $f_0$, $a_0$ properties in a nuclear
medium, where large modifications are expected in view of present
results for the modification of the $\bar{K}$ properties in a 
nuclear medium based on chiral unitary approaches. Such experimental
studies are possible with present facilities like TJNAF and
Spring8/RCNP and they would provide novel tests for our
understanding of the nature of the scalar resonances and about
current ideas on chiral unitary theory, which is emerging as a
powerful tool for the study of meson-meson and meson-baryon interactions.

\vspace{3cm}

Acknowledgments:

We would like to acknowledge useful discussions with A. Titov, T. Nakano
and J.K. Ahn. We are grateful to the COE Professorship program of
Monbusho which enabled E.O. to stay at RCNP to perform the present study.
One of us, E.M., wishes to thank the hospitality of the RCNP of the
University of Osaka, and  acknowledges finantial support from the
Ministerio de Educaci\'on y Cultura. This work is partly
supported by DGICYT contract no. PB 96-07053.

\end{document}